\documentclass[%
groupedaddress,
 amsmath,amssymb,
 aps,
pra,
floatfix,
reprint  % for double-column
]{revtex4-2}

\usepackage{graphicx}% Include figure files
\usepackage{dcolumn}% Align table columns on decimal point
\usepackage{bm}% bold math
\usepackage{hyperref}% add hypertext capabilities
\usepackage[mathlines]{lineno}% Enable numbering of text and display math
\usepackage{subcaption}
\usepackage[dvipsnames]{xcolor}
\usepackage{float}

\newcommand{\changecolor}[1]{#1}

\begin{document}

% \preprint{APS/123-QED}

\title{\textbf{Scattering modeling by a nonlinear slab: exact solution of the full vector problem} 
}% 

\author{Jérémy Itier}
 \email{Contact author: jeremy.itier@fresnel.fr}
\author{Gilles Renversez}
\author{Frédéric Zolla}
 
\affiliation{%
 Aix Marseille Univ, CNRS, Centrale Med, Institut Fresnel, Marseille, France
}%

% \collaboration{CLEO Collaboration}%\noaffiliation

\date{\today}% It is always \today, today,

\begin{abstract}
We investigate the scattering of light by a nonlinear, anisotropic slab under conical incidence and arbitrary polarization, within the framework of Maxwell's equations, where the nonlinearities are described by nonlinear susceptibility tensors. We develop a fully tensorial numerical method, free from standard simplifications such as the undepleted pump approximation or scalar field assumptions, based on an iterative scheme where each step is solved via the finite element method. The two-dimensional problem is reduced to one dimension by exploiting symmetry arguments. Energy considerations are also addressed. \changecolor{Several numerical experiments involving a potassium titanyl phosphate (KTP) slab and a lithium niobate (LiNbO3) photonic crystal are presented, including cases with incident TE and TM waves, as well as a rotation-based study highlighting the anisotropic capabilities of our numerical model.} This work provides a practical and general tool to help the optics research community overcome the limitations of existing models. It may facilitate the design of more advanced experiments to test nonlinear optical theories or improve nonlinear devices.

% \begin{description}
% Secondary publications and information retrieval purposes.
% You may use the \texttt{description} environment to structure your abstract;
% use the optional argument of the \verb+\item+ command to give the category of each item. 
% \end{description}
\end{abstract}

\maketitle

\section{Introduction}
Since the invention of lasers in the early 1960s by Theodore Maiman, the power of light sources has steadily increased, revealing the nonlinear nature of light-matter interactions. From a theoretical perspective, the nonlinearity of the constitutive relations leads, even for a monochromatic source, to a system of coupled, nonlinear partial differential equations of a vector or even tensor nature~\cite{bloembergen_nonlinear_1996, zolla_into_2022}. This takes us out of the well-trodden path of mathematics, where the great theorems (Lax-Milgram, Fredholm's alternative, etc.) guarantee the existence and uniqueness of the solution.

Several simplifications of varying significance are commonly employed, such as the assumption of pump wave non-depletion~\cite{armstrong_interactions_1962, zu_analytical_2022, gladyshev_fast_2024}, which generally leads to the solution of a number of linear differential equations in cascade~\cite{zolla_into_2022-1}. However, these simplifications present at least two major drawbacks: first they no longer hold at high energies~\cite{zolla_into_2022-2}, and second it is difficult to know when these hypotheses are no longer valid without computing the full problem. 
An alternative approach is to tackle only one-dimensional problems. Yet even in this case, the problem remains challenging because the tensor nature of the nonlinear susceptibilities $\chi^{(2)}$ and $\chi^{(3)}$ complicates the task. In particular, obtaining a scalar problem (transverse electric) requires highly restrictive conditions on the crystal. This is one of the reasons why most studies on this subject requires the incident field to illuminate the crystal at normal incidence with transverse polarization, and often consider only a single crystal configuration~\cite{bao_second_1994, jianhua_yuan_exact_2014, szarvas_numerical_2018}.

In this work, we address the nonlinear problem in its full vectorial form, allowing us to move beyond the common assumption of normal incidence and to accommodate arbitrary crystal configurations. Moreover, no assumptions are made regarding the energy of the pump wave, as we solve the complete set of nonlinear equations directly. To the best of our knowledge, such a general method for studying the scattering by a nonlinear, anisotropic slab has not been reported before.

The article is organized as follows: first we recall the general formalism of the system of equations describing the vector electric field when second- and third-order nonlinear processes are considered. \changecolor{Second, for the case of second-harmonic generation, we explicitly derive the equations governing all field components at both the fundamental frequency and the second harmonic, considering all components of \(\chi^{(2)}\). A similar derivation is then carried out for a case including both second- and third- nonlinear processes, with additional details provided in the supplemental document~\cite{supplemental}.} Third, we derive the energy conservation law used to verify the convergence properties of the finite-element implementation of our approach. Fourth, we demonstrate the capabilities of our method through simulations based on realistic configurations using potassium titanyl phosphate (\(\mathrm{KTP}\)) slabs \changecolor{and lithium niobate (\(\mathrm{LiNbO_3}\)) photonic crystals.}

\section{Theoretical model}
Assuming a monochromatic incident field at the frequency \(\omega_I\), we suppose that the total field inside the nonlinear medium can be expanded as \(\mathbf{E}(\mathbf{r},t) = \sum_{p\in\mathbb{Z}^{*}}\mathbf{E}_p(\mathbf{r}) e^{-ip\omega_It}\), with \(\mathbf{E}_p\) and \(\mathbf{E}_{-p}\) \((p \in \mathbb{Z})\) being the complex amplitudes of the field and its conjugate both at frequency \(p\omega_I\). In order to establish our method and to describe its capabilities, we focus only on \(2^{nd}\) and \(3^{rd}\) order nonlinearities, knowing that it can also tackle higher order. Using the framework given in reference \cite{zolla_into_2022}, the set of equations describing the scattering of light when considering only \(2^{nd}\) and \(3^{rd}\) order nonlinearities is:
\begin{equation}
    \begin{array}{c}
         \mathbf{M}_p^{lin}\mathbf{E}_p + \frac{\left(p\omega_I\right)^2}{c^2} \sum_{q\in \mathbb{Z}} \langle\langle  \mathbf{E}_{q}, \mathbf{E}_{p-q} \rangle\rangle \\
         + \frac{\left(p\omega_I\right)^2}{c^2} \sum_{(q,r) \in \mathbb{Z}^2} \langle\langle  \mathbf{E}_{q}, \mathbf{E}_{r}, \mathbf{E}_{p-q-r} \rangle\rangle = -i p \omega \mu_0 \,\mathbf{J}_{p} \delta_{|p|,1}
    \end{array}
    \label{eq:sys_nl}
\end{equation}
The linear part of the Maxwell operator is written \(\mathbf{M}_p^{lin}\) and its definition is given in Eq.~\eqref{eq:Mlin}. The source term is represented by \(\mathbf{J}_p\delta_{|p|,1}\), where \(\delta_{i,j}\) is the Kronecker delta.
The nonlinear parts of the equation are represented using the \(\langle\langle ... \rangle\rangle\) multi-linear operator, involving the susceptibility tensors \(\chi^{(n)}\) of the medium with the field amplitudes by means of contraction (\(:_{(n)}\)) and tensor product (\(\otimes\)), as shown in Eq.~\eqref{eq:nl_operators}.
\begin{equation}
    \mathbf{M}_p^{lin}\mathbf{E}_p  \equiv - \nabla \times \nabla \times \mathbf{E}_p + \frac{(p\omega_I)^2}{c^2}\varepsilon_r(p\omega_I) \mathbf{E}_p
    \label{eq:Mlin}
\end{equation}
\begin{equation}
    \begin{array}{lll}
        \varepsilon_r(p\omega_I) &\equiv 1 + \chi^{(1)}(p\omega_I)\\
        \langle\langle  \mathbf{E}_{p}, \mathbf{E}_{q} \rangle\rangle &\equiv \chi^{(2)}(p\omega_I, q\omega_I) : \mathbf{E}_p \otimes \mathbf{E}_q \\
        \langle\langle  \mathbf{E}_{p}, \mathbf{E}_{q}, \mathbf{E}_{r} \rangle\rangle &\equiv \chi^{(3)}(p\omega_I, q\omega_I, r\omega_I) :_{(3)} \mathbf{E}_p \otimes \mathbf{E}_q \otimes \mathbf{E}_r
    \end{array}
    \label{eq:nl_operators}
\end{equation}

\smallskip

\begin{figure}[ht]
    \centering
    \fbox{\includegraphics[width=0.7\linewidth]{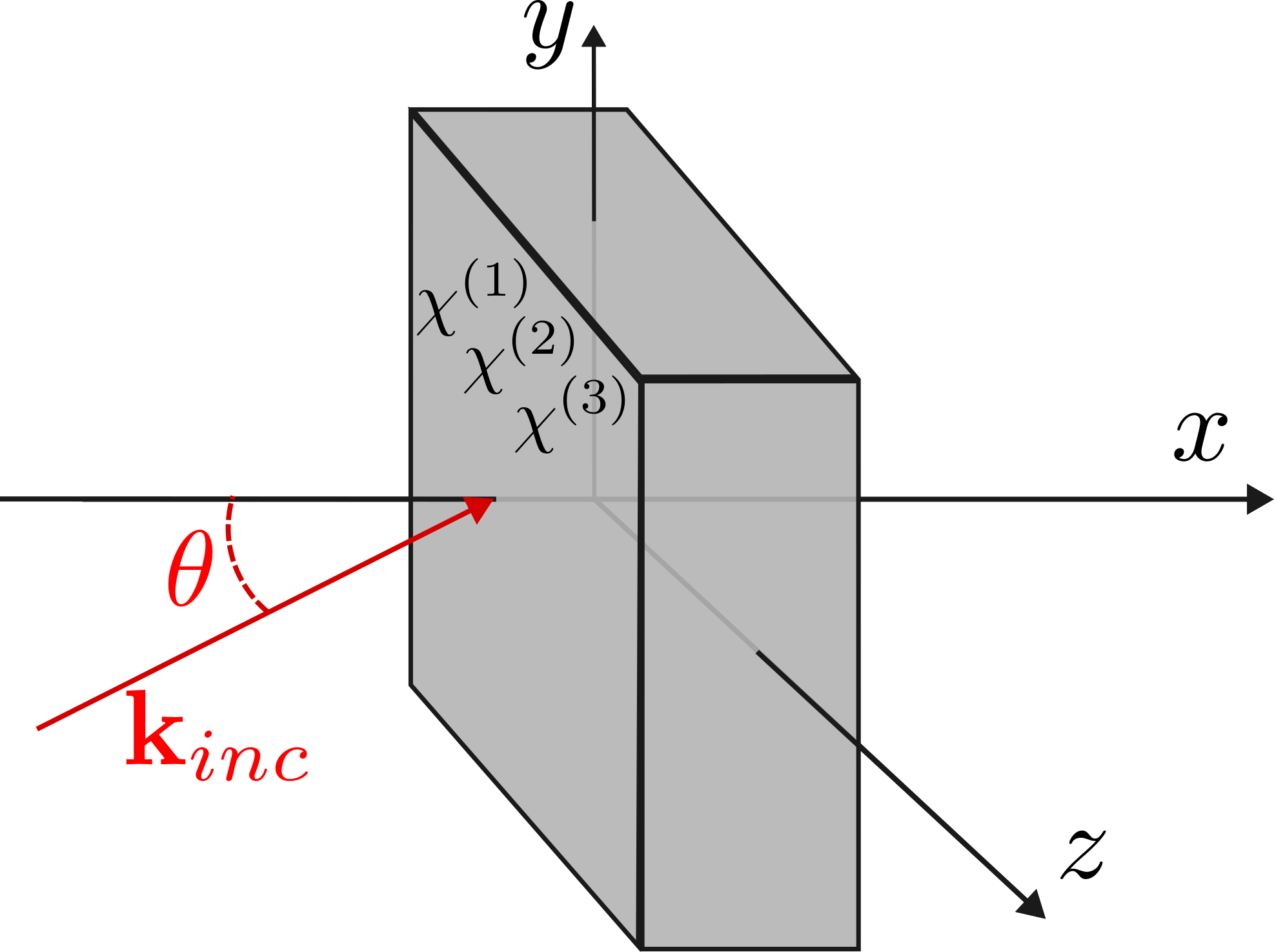}}
    \caption{Schematic view of a nonlinear slab illuminated by a plane wave with its wave vector \(\mathbf{k}_{inc}\) on its left side. \(\mathbf{k}_{inc}\) is within the \((x,y)\) plane.}
    \label{fig:schematic}
\end{figure}

In this work, an infinite nonlinear slab is first considered. The slab is characterized by the susceptibility tensors \(\chi^{(1)}\), \(\chi^{(2)}\), \(\chi^{(3)}\) and in the rest no restriction is required. This is the reason why all the components of the electric field are needed. We place the \(x\) axis perpendicular to the slab and the \(y\) and \(z\) axis parallel to it (see Fig.~\ref{fig:schematic}).

The slab is illuminated by a plane wave in conical incidence. We can choose a basis such that the wave vector \(\mathbf{k}_{inc}\) is contained in the \((\mathbf{e}_x, \mathbf{e}_y)\) plane. In return, the tensors \(\chi^{(1)}\), \(\chi^{(2)}\) and \(\chi^{(3)}\) are a priori arbitrary. We note \(\theta\) the angle between the \(\mathbf{e}_x\) axis and \(\mathbf{k}_{inc}\). Let \(\mathbf{E}_{inc}\) be the complex amplitude of this incident wave:
\begin{equation}
    \mathbf{E}_{inc}(x,y) =  \mathbf{A}_0 e^{i(\alpha x + \beta y)} = \mathbf{A}_0 e^{ik_0(\cos{\theta} \,x + \sin{\theta} \,y)}
    \label{eq:Einc}
\end{equation}
with \(\mathbf{A}_0\) and \(k_0\) being the amplitude and the wave number of the incident wave.

We denote by \(T_y = \frac{2\pi}{k_0 \sin{\theta}}\) the periodicity along the \(y\)-direction of the incident wave. Since both the incident wave and the problem's geometry are \(T_y\)-periodic along \(\mathbf{e}_y\), the resulting field inside the slab must also exhibit \(T_y\)-periodicity. In the special case of normal incidence, the problem becomes independent of \(y\), and the field is accordingly uniform along that direction. Moreover, as shown in the supplemental document \cite{supplemental}, a solution of the form \(\mathbf{E}_p(x,y) = \mathbf{\tilde{E}}_p(x) e^{ip\beta y}\) satisfies the governing equations. In other words, if \(\mathbf{\tilde{E}}_p(x)\) solves the following equation:
\begin{equation}
    \begin{array}{c}
         \mathbf{\tilde{M}}_p^{lin}\mathbf{\tilde{E}}_p  = -i p \omega \mu_0 \,\mathbf{J}_{p} \delta_{|p|,1} - \frac{\left(p\omega_I\right)^2}{c^2} \sum_{q\in \mathbb{Z}} \langle\langle  \mathbf{\tilde{E}}_{q}, \mathbf{\tilde{E}}_{p-q} \rangle\rangle \\
         - \frac{\left(p\omega_I\right)^2}{c^2} \sum_{(q,r) \in \mathbb{Z}^2} \langle\langle  \mathbf{\tilde{E}}_{q}, \mathbf{\tilde{E}}_{r}, \mathbf{\tilde{E}}_{p-q-r} \rangle\rangle
    \end{array}
    \label{eq:sys_nl_ep}
\end{equation}
then \(\mathbf{E}_p(x,y) = \mathbf{\tilde{E}}_p(x) e^{ip\beta y}\) is solution of Eq.~\eqref{eq:sys_nl}, where \(\mathbf{\tilde{M}}^{lin}_p\) is the linear operator defined as:
\begin{equation}
   \mathbf{\tilde{M}}^{lin}_p\left(\mathbf{\tilde{E}}_p(x)\right) \equiv  \mathbf{M}^{lin}_p\left(\mathbf{\tilde{E}}_p(x) e^{ip\beta y}\right) e^{-ip\beta y}  
\end{equation}

\subsection{Second-harmonic generation}
For the first example, we discard the third-order nonlinearities and retain only the first two equations. This leads to the conventional second-harmonic generation (2HG) system commonly found in the literature~\cite{bloembergen_nonlinear_1996}, as shown in Eq.~\eqref{eq:2HG}.
\begin{equation}
    \left\{
        \begin{array}{ll}
            &\mathbf{M}_1^{lin}\mathbf{E}_1 + 2\frac{\omega_I^2}{c^2} \, \langle\langle  \mathbf{E}_{-1}, \mathbf{E}_2 \rangle\rangle = -i\omega\mu_0\,\mathbf{J}_{1} \\
            &\mathbf{M}_2^{lin}\mathbf{E}_2 + \frac{\left(2\omega_I\right)^2}{c^2} \, \langle\langle  \mathbf{E}_1, \mathbf{E}_1 \rangle\rangle = 0
            \label{eq:2HG}
        \end{array}
    \right.
\end{equation}

By making use of Eq.~\eqref{eq:sys_nl_ep} and Eq.~\eqref{eq:2HG}, we find a system of scalar equations, written using Einstein's notation in Sys.~\eqref{eq:nl_scalar}. The derivation in the general case is provided in the supplemental document \cite{supplemental}. \((\tilde{E}_p)_j\), \(\varepsilon_{ij}\) and \(\chi_{ijk}\) respectively stand for the j-th component of \(\mathbf{\tilde{E}}_p\), the (i,j)-th component of the relative permittivity tensor and the (i,j,k)-th component of the susceptibility tensor of \(2^{nd}\) order. The derivatives with respect to \(x\) are denoted by a prime symbol \(\prime\); hence, the first and second derivative of the j-component of the complex amplitude of the field at \(p\omega_I\) are written \((\tilde{E}_p)^{\prime}_j\) and \((\tilde{E}_p)^{\prime\prime}_j\) respectively.

\begin{figure*}[ht]
\begin{subequations}
    \begin{align}
        -i\beta (\tilde{E}_1)^{\prime}_2 - \beta^2 (\tilde{E}_1)_1 + \frac{\omega_I^2}{c^2} \varepsilon_{1j}^{\scriptstyle \omega_I} (\tilde{E}_1)_j + 2\frac{\omega_I^2}{c^2} \chi_{1jk}^{\scriptstyle -\omega_I, 2\omega_I} (\tilde{E}_{-1})_j (\tilde{E}_{2})_k = 0 \\
        (\tilde{E}_1)^{\prime\prime}_2 -i\beta (\tilde{E}_1)^{\prime}_1 + \frac{\omega_I^2}{c^2} \varepsilon_{2j}^{\scriptstyle \omega_I} (\tilde{E}_1)_j + 2\frac{\omega_I^2}{c^2} \chi_{2jk}^{\scriptstyle -\omega_I, 2\omega_I} (\tilde{E}_{-1})_j (\tilde{E}_{2})_k = -i \omega \mu_0 \, (\tilde{J}_{1})_2  \\
        (\tilde{E}_1)^{\prime\prime}_3 - \beta^2 (\tilde{E}_1)_3 + \frac{\omega_I^2}{c^2} \varepsilon_{3j}^{\scriptstyle \omega_I} (\tilde{E}_1)_j + 2\frac{\omega_I^2}{c^2} \chi_{3jk}^{\scriptstyle -\omega_I, 2\omega_I} (\tilde{E}_{-1})_j (\tilde{E}_{2})_k = -i \omega \mu_0 \,(\tilde{J}_{1})_3
    \end{align}
    \begin{align}
        -2i\beta (\tilde{E}_2)^{\prime}_2 - (2\beta)^2 \tilde{E}_2^1 + \frac{\left(2\omega_I\right)^2}{c^2} \varepsilon_{1j}^{\scriptstyle 2\omega_I} (\tilde{E}_2)_j + \frac{\left(2\omega_I\right)^2}{c^2} \chi_{1jk}^{\scriptstyle \omega_I, \omega_I} (\tilde{E}_1)_j (\tilde{E}_{1})_k = 0 \\
        (\tilde{E}_2)^{\prime\prime}_2 -2i\beta (\tilde{E}_2)^{\prime }_1 + \frac{\left(2\omega_I\right)^2}{c^2} \varepsilon_{2j}^{\scriptstyle 2\omega_I} (\tilde{E}_2)_j + \frac{\left(2\omega_I\right)^2}{c^2} \chi_{2jk}^{\scriptstyle \omega_I, \omega_I} (\tilde{E}_1)_j (\tilde{E}_{1})_k = 0 \\
        (\tilde{E}_2)^{\prime\prime}_3 - (2\beta)^2 (\tilde{E}_2)_3 + \frac{\left(2\omega_I\right)^2}{c^2} \varepsilon_{3j}^{\scriptstyle 2\omega_I} (\tilde{E}_2)_j + \frac{\left(2\omega_I\right)^2}{c^2} \chi_{3jk}^{\scriptstyle \omega_I, \omega_I} (\tilde{E}_1)_j (\tilde{E}_{1})_k = 0
    \end{align} \label{eq:nl_scalar}
\end{subequations}
\end{figure*}

This leads to a system of six coupled nonlinear ordinary differential equations, dependent only on the \(x\)-variable. Because the problem is one-dimensional, numerical experiments can be performed over large slab lengths with high accuracy. The remaining challenge lies in solving these nonlinear equations.

\subsection{Third-order nonlinear processes}

\changecolor{To further demonstrate the capabilities of our method, we also consider a more complex example involving both second- and third-order nonlinear processes with three harmonics. By including all the terms of Eq.~\eqref{eq:sys_nl} and restricting the analysis to the first three harmonics, we obtain the system given in Sys.~\eqref{eq:3HG}. This system captures several nonlinear effects: second-order processes such as second-harmonic generation \((\omega+\omega\rightarrow2\omega)\) and sum-frequency generation \((2\omega+\omega\rightarrow 3\omega)\); third-harmonic generation \((\omega+\omega+\omega\rightarrow3\omega)\); and both self- and cross-phase modulation resulting from the interactions among different harmonics.}
\small
\begin{equation} 
    \left\{
        \begin{array}{ll}
            &\changecolor{\mathbf{M}_1^{lin}\mathbf{E}_1 + \frac{\omega_I^2}{c^2} \Bigl( 2\langle\langle  \mathbf{E}_{-1}, \mathbf{E}_2 \rangle\rangle + 2\langle\langle  \mathbf{E}_{-2}, \mathbf{E}_3 \rangle\rangle} \\
            &\changecolor{+ 3 \langle\langle \mathbf{E}_{-1}, \mathbf{E}_{1}, \mathbf{E}_{1} \rangle\rangle + 6 \langle\langle \mathbf{E}_{-2}, \mathbf{E}_{1}, \mathbf{E}_{2} \rangle\rangle + 6 \langle\langle \mathbf{E}_{-3}, \mathbf{E}_{1}, \mathbf{E}_{3} \rangle\rangle} \\
            &\changecolor{+ 3 \langle\langle \mathbf{E}_{-1}, \mathbf{E}_{-1}, \mathbf{E}_{3} \rangle\rangle + 3 \langle\langle \mathbf{E}_{2}, \mathbf{E}_{2}, \mathbf{E}_{-3} \rangle\rangle \Bigr)}\\
            &\changecolor{= -i\omega\mu_0\,\mathbf{J}_{1}} \\
            &\changecolor{\mathbf{M}_2^{lin}\mathbf{E}_2 + \frac{(2\omega_I)^2}{c^2} \Bigl( 2\langle\langle  \mathbf{E}_{-1}, \mathbf{E}_3 \rangle\rangle + \langle\langle  \mathbf{E}_{1}, \mathbf{E}_1 \rangle\rangle} \\
            &\changecolor{+ 6 \langle\langle \mathbf{E}_{-1}, \mathbf{E}_{2}, \mathbf{E}_{1} \rangle\rangle + 3 \langle\langle \mathbf{E}_{-2}, \mathbf{E}_{2}, \mathbf{E}_{2} \rangle\rangle + 6 \langle\langle \mathbf{E}_{-3}, \mathbf{E}_{2}, \mathbf{E}_{3} \rangle\rangle} \\
            &\changecolor{+ 6 \langle\langle \mathbf{E}_{-2}, \mathbf{E}_{1}, \mathbf{E}_{3} \rangle\rangle \Bigr)}\\
            &\changecolor{= 0} \\
            &\changecolor{\mathbf{M}_3^{lin}\mathbf{E}_3 + \frac{(3\omega_I)^2}{c^2} \Bigl( 2\langle\langle  \mathbf{E}_{1}, \mathbf{E}_2 \rangle\rangle} \\
            &\changecolor{+ 6 \langle\langle \mathbf{E}_{-1}, \mathbf{E}_{3}, \mathbf{E}_{1} \rangle\rangle + 6 \langle\langle \mathbf{E}_{-2}, \mathbf{E}_{3}, \mathbf{E}_{2} \rangle\rangle + 3 \langle\langle \mathbf{E}_{-3}, \mathbf{E}_{2}, \mathbf{E}_{3} \rangle\rangle} \\
            &\changecolor{+ 3 \langle\langle \mathbf{E}_{-1}, \mathbf{E}_{2}, \mathbf{E}_{2} \rangle\rangle + \langle\langle \mathbf{E}_{1}, \mathbf{E}_{1}, \mathbf{E}_{1} \rangle\rangle \Bigr)}\\
            &\changecolor{= 0}
            \label{eq:3HG}
        \end{array}
    \right.
\end{equation}
\normalsize

\changecolor{Following the procedure used for the second-harmonic generation case, we derive the scalar system describing the one-dimensional problem by making use of Eq.~\eqref{eq:sys_nl_ep} and Sys.~\eqref{eq:3HG}, as detailed in the supplemental document~\cite{supplemental}.}

\section{Numerical computation}
Solving partial differential linear equations is quite straightforward, thanks to numerous established numerical techniques such as the finite element method (FEM) and the finite difference method (FDM). However, when dealing with coupled nonlinear equations --- such as system~(\ref{eq:nl_scalar}) --- these tools generally become inadequate and require specific adaptations. We opted for the finite element method, as it naturally extends to more complex geometries in two and three dimensions.

The core idea of the method is to transform the system of coupled partial differential equations into a matrix system of the form \(\mathbf{A}\mathbf{x}=\mathbf{b}(\mathbf{x})\). Since the right-hand side depends on \(\mathbf{x}\), the matrix equation is nonlinear with respect to \(\mathbf{x}\) and cannot be solved using standard linear solvers. One approach is to linearize the system using Picard iterations (also known as the fixed-point method), which we have successfully applied in the context of nonlinear eigenvalue problems for the study of waveguides \cite{elsawy_study_2017, elsawy_exact_2018}. The method consists in solving \(\mathbf{A}\mathbf{x_i}=\mathbf{b}(\mathbf{x_{i-1}})\) at each iteration, replacing the nonlinear term with the solution of the previous step. Existence and uniqueness results were derived in \cite{bao_second_1994} for the specific case of 2HG. Although alternative algorithms with faster convergence exist, such as the Newton-Raphson method, they are not discussed in this article. More details are provided in the supplemental document \cite{supplemental}.

The FEM simulations were performed using \textit{Gmsh} \cite{geuzaine_gmsh_2009} and \textit{GetDP} \cite{dular_general_1998}, two open-source software packages. Several challenges had to be addressed. First, due to the nonlinearities, the conventional approach using a diffracted field to simulate an incident plane wave from infinity is not applicable. Instead, a virtual antenna technique was employed, consisting of applying a tailored surface current to generate the desired wave, as described in \cite{zolla_virtual_2006}. Outgoing wave conditions were applied at the domain boundaries as they are exact in one-dimensional problems.

Another challenge arises from the fact that the normal component of the electric field is expected to be discontinuous across interfaces. However, the use of nodal elements in the finite element method inherently enforce continuity. In two dimensions, this issue could be overcome using edge elements, but in our one-dimensional setting, a different approach is required. To address this, the normal components of the electric field are split into three distinct sets of unknowns, each defined within a specific region of the domain: substrate, slab, and superstrate. These unknowns are coupled only through the appropriate boundary conditions, avoiding any artificial continuity constraints. This approach can be regarded as a simplified form of the \textit{discontinuous Galerkin method}.~\cite{cockburn_discontinuous_2000}.

\section{Energy study}
In order to validate our approach and test the convergence properties of the FEM with respect to the mesh size, we develop a specific energy conservation rule from basic principles.
It allows to quantify the energy exchange between the harmonics themselves and the material.
We start with the energy conservation equation:
\begin{equation}
    -\partial_t w = \nabla \cdot \boldsymbol{\Pi} + \mathbf{E} \cdot \mathbf{J}
    \label{eq:energy-balance-init}
\end{equation}
with \(\partial_t w = \partial_t w_m + \partial_t w_e\) being the time derivative of the total electromagnetic energy made of the electric term \(w_e\) and the magnetic one \(w_m\), and \(\boldsymbol{\Pi}\) being the Poynting vector. The term \(\partial_t w_e\) can be decomposed by following the order of the nonlinearity: \(\partial_t w_e = \sum_{k \in \mathbb{Z}} \partial_t w_e^{(k)}\).

A general expression of the time average \(<\partial_t w_e^{(k)}>\) is given in \cite{zolla_into_2022-1}, and are shown in Eq.~\eqref{eq:energy_terms_general}.
\begin{equation}
    \begin{array}{llll}
        &< \partial_t w_e^{(0)} > = 0 \\
        &< \partial_t w_e^{(1)} > = -i\epsilon_0\omega \sum_{q \in \mathbb{Z}}  q \mathbf{E}_{-q} \cdot \langle\langle \mathbf{E}_{q} \rangle\rangle \\
        &< \partial_t w_e^{(2)} > = -i\epsilon_0\omega \sum_{(q,r) \in \mathbb{Z}^2} r \mathbf{E}_{-r} \cdot \langle\langle \mathbf{E}_{q}, \mathbf{E}_{r-q} \rangle\rangle \\
        &< \partial_t w_e^{(3)} > = -i\epsilon_0\omega \sum_{(p,q,r) \in \mathbb{Z}^3} r \mathbf{E}_{-r} \cdot \langle\langle \mathbf{E}_{p}, \mathbf{E}_{q}, \mathbf{E}_{r-p-q} \rangle\rangle
    \end{array} \label{eq:energy_terms_general}
\end{equation}

\changecolor{By limiting the number of harmonics, the general expression can be simplified. For instance, in the case of second-harmonic generation (SHG), it reduces to:}
\begin{equation}
    \hspace{-0.6cm}
    \begin{array}{llll}
        &<\partial_t w_e^{(0)}> = 0 \\
        &<\partial_t w_e^{(1)}> = 2\epsilon_0 \omega_I \mathfrak{Im}\{\mathbf{E}_{-1} \cdot \langle\langle \mathbf{E}_{1} \rangle\rangle + 2\mathbf{E}_{-2} \cdot \langle\langle \mathbf{E}_{2} \rangle\rangle \} \\
        &<\partial_t w_e^{(2)}> = 4\epsilon_0 \omega_I \mathfrak{Im}\{\mathbf{E}_{-1} \cdot \langle\langle \mathbf{E}_{-1}, \mathbf{E}_{2} \rangle\rangle + \mathbf{E}_{-2} \cdot \langle\langle \mathbf{E}_{1}, \mathbf{E}_{1} \rangle\rangle \} \\
        &<\partial_t w_e^{(3)}> = 0
    \end{array} \label{eq:energy_terms_2HG}
\end{equation}

Assuming there is no volume current in the material, the time-averaged energy balance Eq.~\eqref{eq:energy-balance-init} over a domain \(\Omega\) can be written as:
\begin{equation}
    -\iiint_\Omega <\partial_t w> \mathrm{d}^3\tau = \iint_{\partial\Omega} <\boldsymbol{\Pi}> . \, \mathrm{d}^2\mathbf{s}
    \label{eq:energy_balance}
\end{equation}

In the case of an infinite slab, we can rewrite the Poynting surface integral with reflection and transmission coefficients. It saves us from computing the spatial derivatives of the electric field.
\begin{equation}
    \iint_{\partial\Omega} <\boldsymbol{\Pi}> . \, \mathrm{d}^2\mathbf{s} = \frac{2n\cos{\theta}|\mathbf{A}_0|^2}{c\mu_0}\Biggl(\sum_p (R_p + T_p) -1\Biggr)
    \label{eq:poynting_to_coeff}
\end{equation}
\(\mathbf{A}_0\) is the amplitude of the incident wave, \(n\) the index of the substrate and superstrate, \(R_p\) and \(T_p\) are, respectively, the reflection and transmission coefficients at \(p\omega_I\). Using equations \eqref{eq:energy_balance} and \eqref{eq:poynting_to_coeff}, we can compute the energy balance of our problem and assess the validity of our models. A more detailed derivation is provided in the supplemental document \cite{supplemental}.

\section{Numerical experiments}
\subsection{KTP simulations}

To illustrate the capabilities of our method, we consider a slab of KTP, a material with an orthorhombic crystal structure belonging to the \(mm2\) point group symmetry \cite{fan_second_1987}. In a basis \((X, Y, Z)\) aligned with the crystal lattice --- distinct, a priori, from the one associated with the slab --- these symmetries correspond to a two-fold rotation around the \(Z\)-axis and two mirror planes perpendicular to the \(X\) and \(Y\) axes. Such symmetry imposes constraints on the susceptibility tensors \(\chi^{(1)}\) and \(\chi^{(2)}\), reducing the number of their independent components.

For KTP, assuming that the intrinsic symmetry is verified, \(\chi^{(1)}\) and \(\chi^{(2)}\) are reduced to 3 and 5 independent components respectively \cite{boyd_nonlinear_2020}. Linear parameters from reference~\cite{fan_second_1987} are provided in Table~\ref{tab:linear_parameters_KTP}. Regarding the second-order susceptibility tensor, some measurements have been reported in the past \cite{vanherzeele_magnitude_1992, boulanger_relative_1994, shoji_absolute_1997}, but only for \(\chi^{(2)}(\omega_I, \omega_I)\), and thus \(\chi^{(2)}(-\omega_I, 2\omega_I)\) must be inferred from the available experimental data. The more recent and comprehensive dataset has been retained for this purpose \cite{shoji_absolute_1997}.

One option is to assume a non-dispersive second-order susceptibility, i.e. \(\chi^{(2)}(-\omega_I, 2\omega_I)=\chi^{(2)}(\omega_I, \omega_I)\). However, since the measured \(\chi^{(2)}(\omega_I, \omega_I)\) reported in \cite{shoji_absolute_1997} does not meet the usual Kleinman symmetry criterion \cite{boyd_nonlinear_2020} --- namely, \(\chi^{(2)}_{ijk}(\omega_I, \omega_I) \neq \chi^{(2)}_{jki}(\omega_I, \omega_I)\) --- this assumption would imply that the material behaves as both active and passive depending on its orientation, which appears unrealistic.
\emph{Instead, we adopt an alternative approach by constructing \(\chi^{(2)}(-\omega_I, 2\omega_I)\) so that the full permutation symmetry is satisfied, i.e., \(\chi^{(2)}_{ijk}(-\omega_I, 2\omega_I) = \chi^{(2)}_{jki}(\omega_I, \omega_I)\), ensuring that the KTP slab remains lossless~\cite{boyd_nonlinear_2020, godard_optique_2009}} (see Sys.~\eqref{eq:susceptibility_tensor}). In practice, both approaches lead to very similar numerical values, since the susceptibility tensor from \cite{shoji_absolute_1997} nearly satisfies Kleinman symmetry, with \(d_{15} = d_{31}\) and \(d_{24} \simeq d_{32}\), as shown in Table~\ref{tab:nl_parameters_KTP}.

\begin{subequations}
\begin{align}
    (\chi_{1jk}^{\scriptstyle \omega_I, \omega_I}) &= 
    \begin{pmatrix}
        0 & 0 & 2d_{15} \\
        0 & 0 & 0 \\
        2d_{15} & 0 & 0
    \end{pmatrix}_{(X, Y, Z)} \nonumber\\
    (\chi_{2jk}^{\scriptstyle \omega_I, \omega_I}) &= 
    \begin{pmatrix}
        0 & 0 & 0 \\
        0 & 0 & 2d_{24} \\
        0 & 2d_{24} & 0
    \end{pmatrix}_{(X, Y, Z)}\\
    (\chi_{3jk}^{\scriptstyle \omega_I, \omega_I}) &= 
    \begin{pmatrix}
        2d_{31} & 0 & 0 \\
        0 & 2d_{32} & 0 \\
        0 & 0 & 2d_{33}
    \end{pmatrix}_{(X, Y, Z)} \nonumber
\end{align} 

\begin{align}
    (\chi_{1jk}^{\scriptstyle -\omega_I, 2\omega_I}) &= 
    \begin{pmatrix}
        0 & 0 & 2d_{31} \\
        0 & 0 & 0 \\
        2d_{15} & 0 & 0
    \end{pmatrix}_{(X, Y, Z)} \nonumber\\
    (\chi_{2jk}^{\scriptstyle -\omega_I, 2\omega_I}) &= 
    \begin{pmatrix}
        0 & 0 & 0 \\
        0 & 0 & 2d_{32} \\
        0 & 2d_{24} & 0
    \end{pmatrix}_{(X, Y, Z)}\\
    (\chi_{3jk}^{\scriptstyle -\omega_I, 2\omega_I}) &= 
    \begin{pmatrix}
        2d_{15} & 0 & 0 \\
        0 & 2d_{24} & 0 \\
        0 & 0 & 2d_{33}
    \end{pmatrix}_{(X, Y, Z)} \nonumber 
\end{align} \label{eq:susceptibility_tensor}
\end{subequations}

\begin{table}[htbp]
    \centering
    \caption{\bf Linear parameters of KTP \cite{fan_second_1987}}
    \begin{tabular}{cccc}
        \hline
        \(\lambda \,(\mathrm{nm})\) & \(n_x\) & \(n_y\) & \(n_z\) \\
        \hline
        $1064$ & $1.7381$ & $1.7458$ & $1.8302$ \\
        $532$ & $1.7780$ & $1.7887$ & $1.8888$ \\
        \hline
    \end{tabular}
    \label{tab:linear_parameters_KTP}
\end{table}

\begin{table}[htbp]
    \centering
    \caption{\bf Nonlinear parameters of KTP, at \(\mathbf{\lambda_0 = \,1064nm}\), in [m/V] \cite{shoji_absolute_1997}}
    \begin{tabular}{cccccc}
        \hline
        \(d_{15}\) & \(d_{24}\) & \(d_{31}\) & \(d_{32}\) & \(d_{33}\) \\
        \hline
        $3.7 \,.10^{-12} $ & $1.9 \,.10^{-12}$ & $3.7 \,.10^{-12}$ & $2.2 \,.10^{-12}$ & $14.6 \,.10^{-12}$ \\
        \hline
    \end{tabular}
    \label{tab:nl_parameters_KTP}
\end{table}

We begin by studying the cases of incident TE and TM waves. Since the material is anisotropic, an incident wave does not necessarily maintain its polarization as it propagates through the slab. Therefore, the usual scalar model is valid only under very specific crystal configurations. We investigate the conditions under which an incident TE wave preserves its initial polarization during propagation. To ensure that the electric field remains polarized along \(\mathbf{e}_z\), the electric displacement field must also lie along \(\mathbf{e}_z\). Given that \(e_p^x=e_p^y=0\) in this case, the equations in Sys.~\eqref{eq:nl_operators} reduce to:
\begin{subequations}
    \begin{align}
        &\chi^{(1)} \,.\, \mathbf{E}_p = \chi^{(1)}_{ij} \, E_p^j \, \mathbf{e}_i = \chi^{(1)}_{i3} \, E_p^z \, \mathbf{e}_i \\
        &\chi^{(2)} : \mathbf{E}_p \otimes \mathbf{E}_q = \chi^{(2)}_{ijk} \, E_p^j \, E_q^k \, \mathbf{e}_i = \chi^{(2)}_{i33} \, E_p^z \, E_q^z \, \mathbf{e}_i
    \end{align} \label{eq:tensor_explicit_te}
\end{subequations}
The right-hand side terms in Eq.~\eqref{eq:tensor_explicit_te} must be colinear with \(\mathbf{e}_z\), which implies that the components \(\chi^{(1)}_{13}\), \(\chi^{(1)}_{23}\), \(\chi^{(2)}_{133}\) and \(\chi^{(2)}_{233}\) must vanish. While this condition holds for KTP susceptibilities expressed in the original basis (Sys.~\eqref{eq:susceptibility_tensor}), any rotation of the slab breaks it, thereby requiring a fully vectorial method to accurately model the problem. \emph{In any case, the incident TM case also requires a fully vectorial method, as the nonlinearities prevent the problem from being expressed solely in terms of the magnetic field.}

\smallskip

\begin{figure*}[ht]
    \centering
    \begin{subfigure}{0.49\linewidth}
        \centering
        \includegraphics[width=\linewidth]{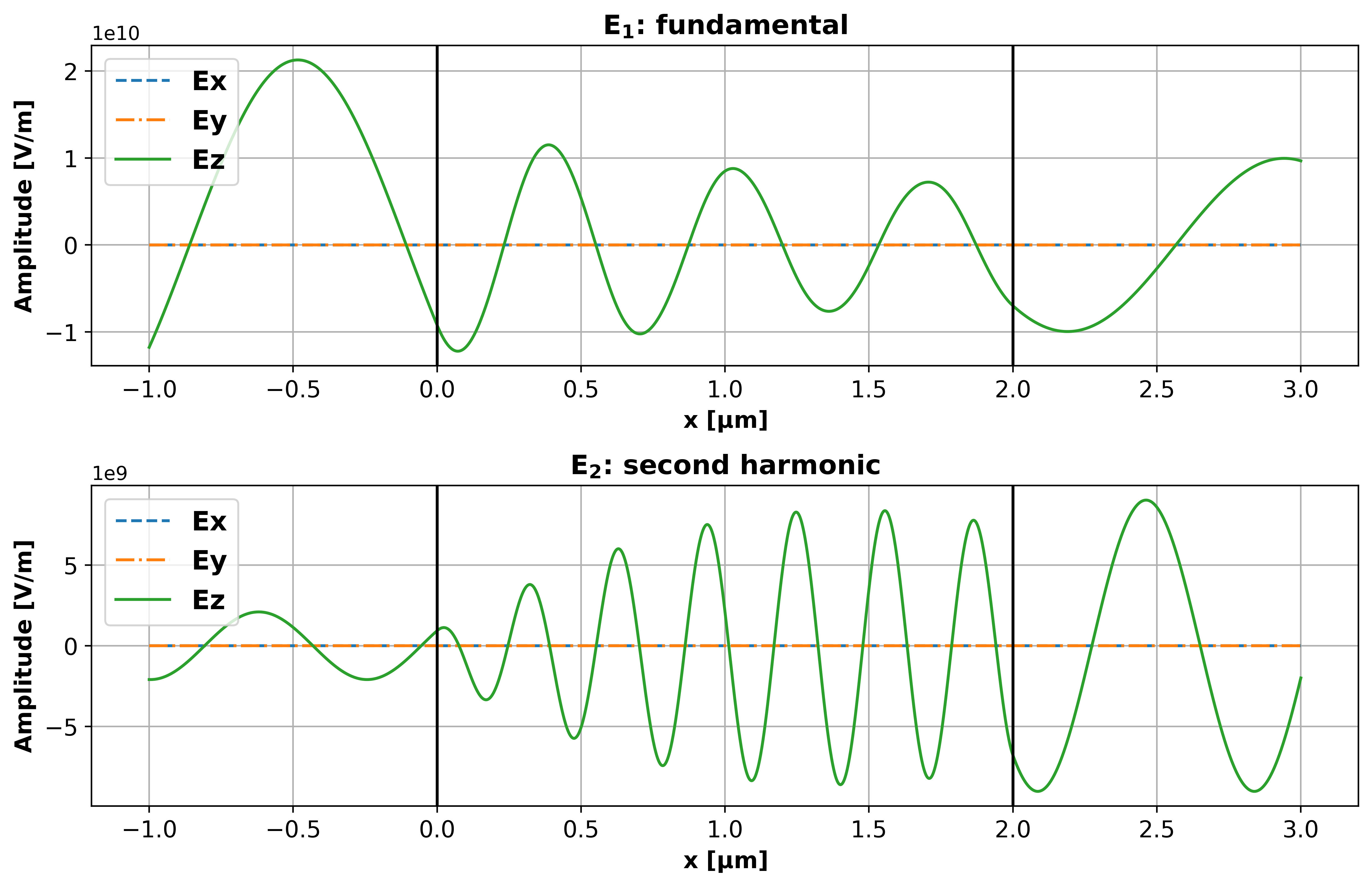}
        \caption{Incident TE-polarized wave.}
        \label{fig:2HG_TE}
    \end{subfigure}
    \begin{subfigure}{0.49\linewidth}
        \centering
        \includegraphics[width=\linewidth]{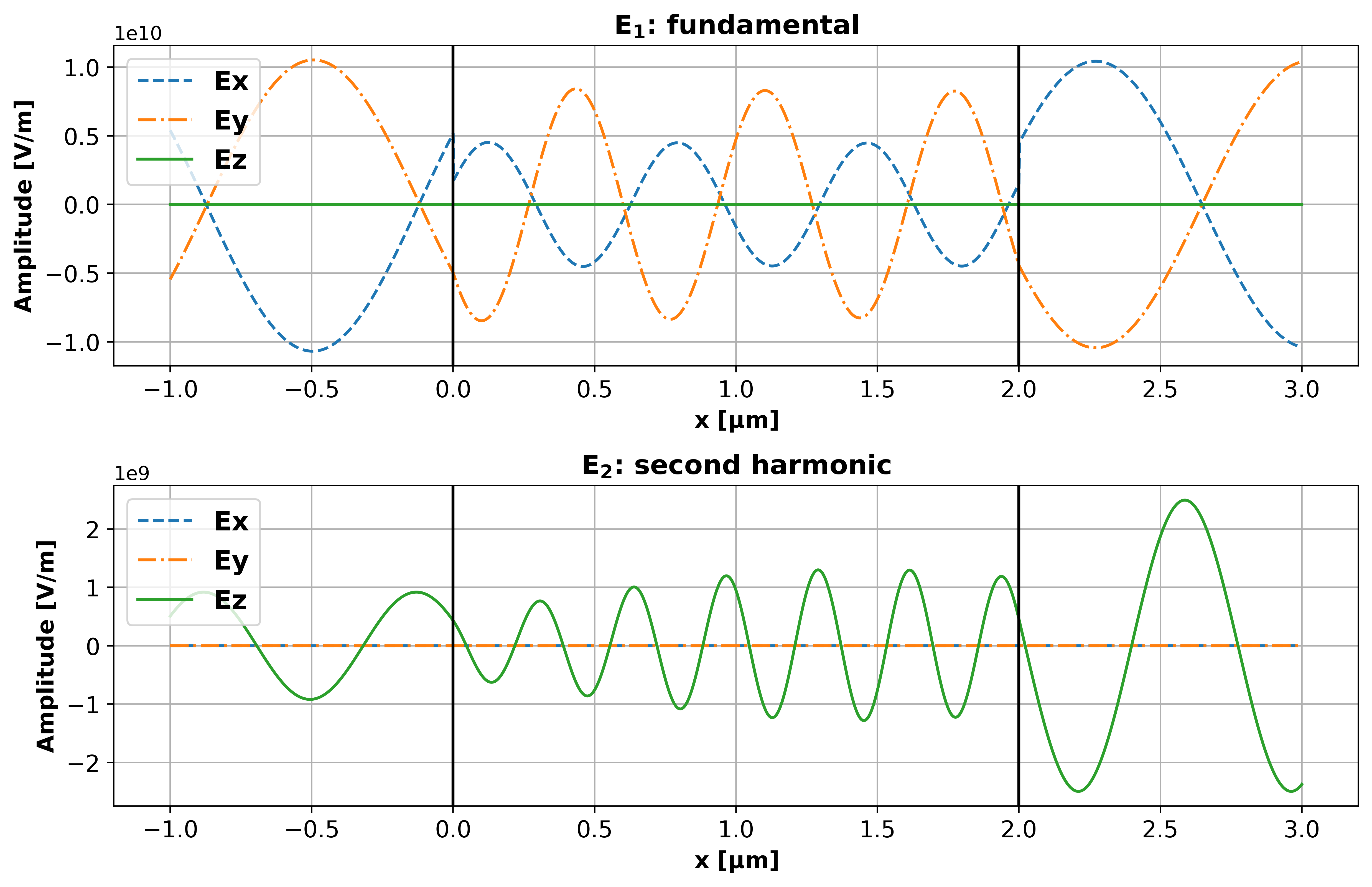}
        \caption{Incident TM-polarized wave.}
        \label{fig:2HG_TM}
    \end{subfigure}
    \caption{Nonlinear scattering from a KTP slab under TE- and TM-polarized illumination. The real parts of the fundamental field \(\mathbf{E}_1\) and the second-harmonic \(\mathbf{E}_2\) are plotted along the \(x\)-axis ($y$$=$$0$), for an incident plane wave from the left with an amplitude \(A_0 = 1,5\times10^{10} V/m\) and an incidence angle \(\theta=\pi/4 \ rad\). The two solid black vertical bars indicate the slab interfaces.}
    \label{fig:field_2HG}
\end{figure*}

A plot of the electric field for an incident TE plane wave is shown in Fig.~\ref{fig:2HG_TE}. The results are consistent with those previously obtained within the scalar framework \cite{jianhua_yuan_exact_2014}. An increasing second harmonic is generated as the wave passes through the slab while the fundamental field decreases, indicating an energy transfer between the two waves. This clearly demonstrates that the commonly used assumption of non-depletion of the pump wave does not hold here. It is also worth noting the effect of phase mismatch on the generation of the second harmonic beyond \(0.5\mu m\): the propagating part of the harmonic interferes destructively with the newly generated part, resulting in a decrease in its amplitude.

In the TM case (Fig.~\ref{fig:2HG_TM}), no harmonic is generated along the \(x\) and \(y\) axes, as the nonlinear coefficients associated with these directions vanish. However, a second harmonic is generated along the \(z\)-axis, due to the presence of the \(d_{31}\) and \(d_{32}\) coefficients coupling the TE and TM waves. The amplitude of the generated harmonic is lower than in the TE case because the dominant nonlinear coefficient responsible for harmonic generation in the TE case (\(d_{33}\)) is nearly five times larger than those involved in the TM case (\(d_{31}\) and \(d_{32}\)). As expected, the normal components of the field are discontinuous at the interface.

\begin{figure*}[ht]
    \centering
    \begin{subfigure}{0.49\linewidth}
        \centering
        \includegraphics[width=\linewidth]{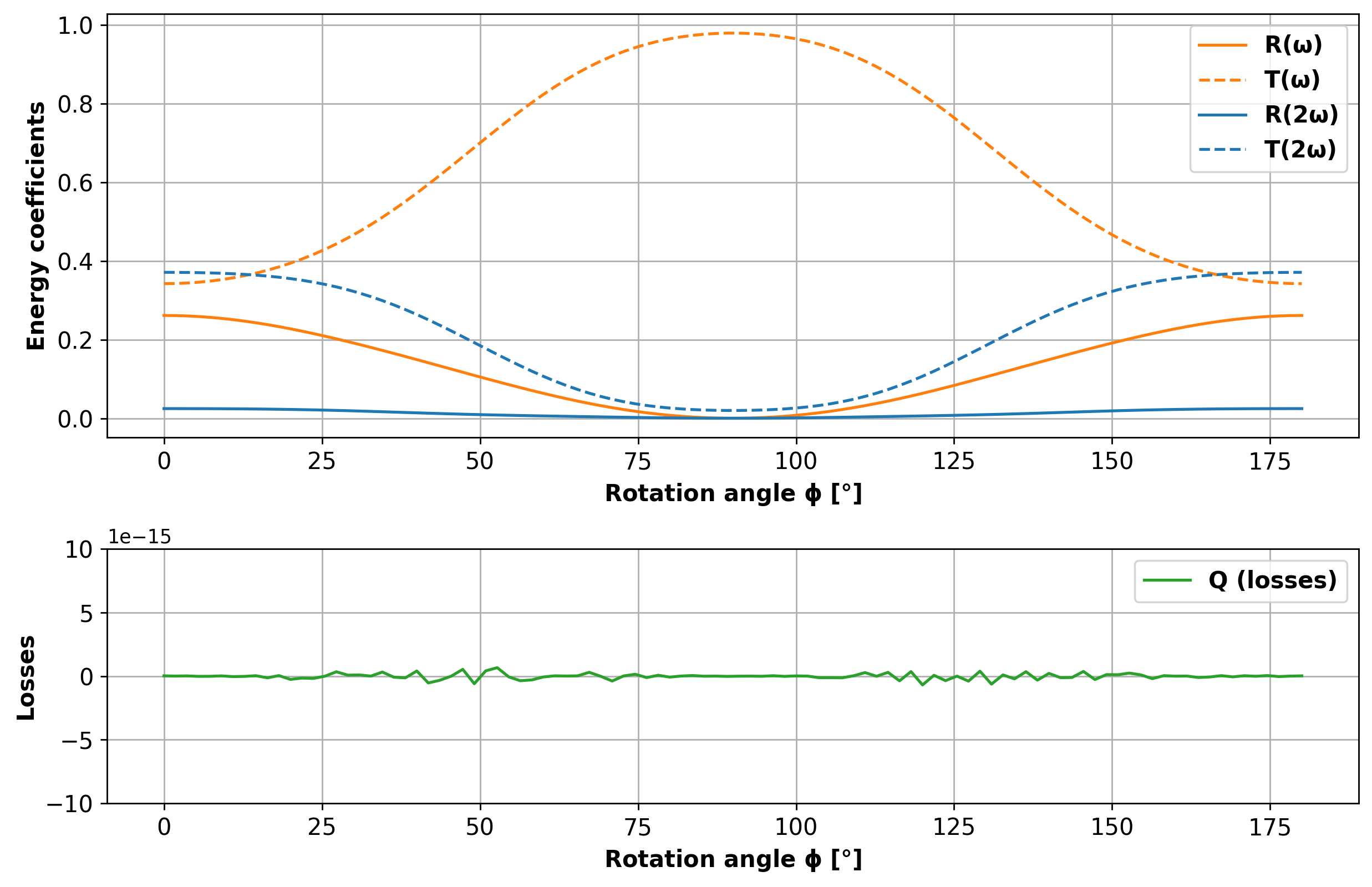}
        \caption{Energy coefficients as a function of the angle of rotation of the crystal \(\phi\), for a TE incident plane wave \((\gamma=\pi/2)\).}
        \label{fig:RTcoefficients_phi}
    \end{subfigure}
    \begin{subfigure}{0.49\linewidth}
        \centering
        \includegraphics[width=\linewidth]{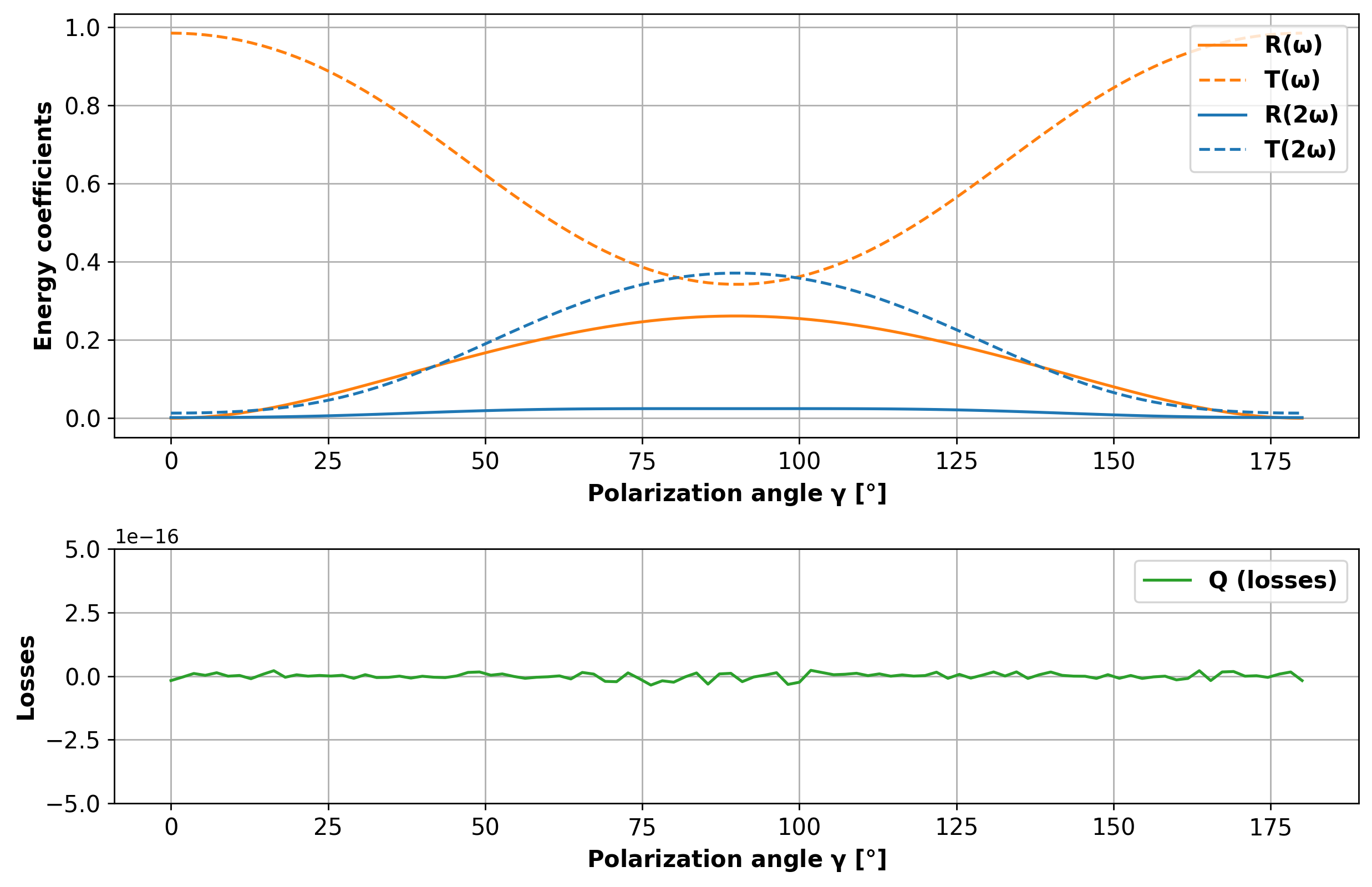}
        \caption{\changecolor{Energy coefficients as a function of the polarization angle~\(\gamma\).}}
        \label{fig:RTcoefficients_gamma}
    \end{subfigure}
    \caption{Reflection (\(R_1\), \(R_2\)) and transmission (\(T_1\), \(T_2\)) coefficients of the slab at \(\omega_I\) and \(2\omega_I\), as well as normalized losses \((Q)\), are shown as a function of the crystal rotation angle \(\phi\) around the \(x\)-axis (a) \changecolor{and of the polarization angle \(\gamma\) around the incident wave vector \(\mathbf{k}_{inc}\) (b)}. The simulations were carried out for a plane wave with an amplitude \(A_0 = 10^{10} V/m\) and an incidence angle \(\theta=\pi/4\), using a KTP slab of thickness \(2 \, \mathrm{\mu m}\). Energy conservation is verified within \(6 \times10^{-7}\%\) for all the studied angles.}
    \label{fig:RTcoefficients}
\end{figure*}

% \begin{figure}[ht]
%     \centering
%     \includegraphics[width=\linewidth]{figures/PlotEnergyCoefficentAndLosses_phi.png}
%     \caption{Reflection (\(R_1\), \(R_2\)) and transmission (\(T_1\), \(T_2\)) coefficients of the slab respectively at \(\omega_I\) and \(2\omega_I\), and normalized losses \((Q)\) are shown as a function of the angle of rotation of the crystal \(\phi\) around the \(x\)-axis. The simulation was carried out for a TE incident plane wave with amplitude \(A_0 = 10^{10} V/m\) and angle of incidence \(\theta=\pi/4\), using a KTP slab of thickness \(2.4\ \, \mathrm{\mu m}\). Energy conservation is satisfied with an accuracy better than \(6 \times10^{-7}\%\) for all the studied angles.}
%     \label{fig:RTcoefficients}
% \end{figure}

\smallskip

The model we present is highly general and suitable for experimentally feasible numerical experiments. For instance, one can consider a conical incident wave and rotate the KTP crystal around the \(x\)-axis while measuring the reflection and transmission coefficients. This is equivalent to rotating the wave vector around the \(x\)-axis. The results of this numerical experiment are shown in Fig.~\ref{fig:RTcoefficients_phi}. A significant energy conversion of approximately 30\% is observed at 0°, which gradually decreases as the crystal is rotated, reaching around 1\% at 90°. This behavior is explained by the vanishing of the main contribution involved in second-harmonic generation in the TE case, \(\chi^{(2)}_{333}\), after a 90° tensor rotation around the \(x\)-axis (see the supplemental document \cite{supplemental}). As expected, the losses remain nearly zero across all angles, consistent with the enforced full permutation symmetry.

\changecolor{A similar experiment can be performed by measuring the energy coefficients as a function of the polarization angle \(\gamma\) defined around the incident wave vector \(\mathbf{k}_{inc}\), with \(\gamma=0\) and \(\gamma=\pi/2\) corresponding to TM and TE polarizations, respectively. The definition of the incident wave amplitude is given in Eq.~\eqref{eq:A0}. The results are shown in Fig.~\ref{fig:RTcoefficients_gamma}. We observe a pronounced dependence on the incident polarization, highlighting its crucial role. In particular,  the energy conversion is nearly zero for TM polarization, while it is significant for TE polarization. As in the previous case, this behavior is explained by the fact that a TM plane wave (\(\gamma=0\)) does not interact with the dominant nonlinear coefficient \(\chi^{(2)}_{333}\).}
\begin{equation}
    \changecolor{\mathbf{A}_0=A_0
    \begin{pmatrix}
        -\cos{\gamma \sin{\theta}} \\
        \cos{\gamma \cos{\theta}} \\
        \sin{\gamma}
    \end{pmatrix}} \label{eq:A0}
\end{equation}

% \pagebreak
\subsection{LiNbO3 case}

\changecolor{To illustrate the ability of the method to also handle third-order nonlinear processes, we consider a slab and a photonic crystal made of \(\mathrm{LiNbO_3}\). The material parameters are given in Tables~\ref{tab:linear_parameters_LiNbO3} and~\ref{tab:nl_parameters_LiNbO3}. As accurate measurements of nonlinear parameters are scarce in the literature, the second- and third-order nonlinear susceptibility tensors are assumed to be non-dispersive and to satisfy Kleinman symmetry. We use \(d_{33}=13.6 \,\, \mathrm{pm/V}\) to compare our results with Ref.~\cite{szarvas_numerical_2018}, although this value differs from the expected one, likely due to a missing factor of two; the correct one should be \(d_{33}=27.2\, \mathrm{pm/V}\), as reported in the original source \cite{nikogosyan_nonlinear_2005}. Third-order nonlinear parameters are extrapolated from Ref.~\cite{kulagin_analysis_2006}. However, since no value has been reported in the literature for \(\chi_{1123}\), we set it equal to \(\chi_{3333}\).}

\begin{table}[htbp]
    \centering
    \caption{\changecolor{\bf Linear parameters of LiNbO3 \cite{zelmon_infrared_1997}}}
    \begin{tabular}{cccc}
        \hline
        \(\lambda \,(\mathrm{nm})\) & \(n_o\) & \(n_e\) \\
        \hline
        $1064$ & $2.2321$ & $2.1555$ \\
        $532$ & $2.3232$ & $2.2336$ \\
        $355$ & $2.4393$ & $2.3321$ \\
        \hline
    \end{tabular}
    \label{tab:linear_parameters_LiNbO3}
\end{table}
\begin{table}[htbp]
    \centering
    \caption{\changecolor{\bf Nonlinear parameters of LiNbO3, at \(\mathbf{\lambda_0 =1064nm}\), in [m/V] and [m\(²\)/V\(^2\)] \cite{nikogosyan_nonlinear_2005, kulagin_analysis_2006}}}
    \begin{tabular}{ccc}
        \hline
        \(d_{22}\) & \(d_{31}\) & \(d_{33}\) \\
        \hline
        $2.1 \,.10^{-12} $ & $-4.35 \,.10^{-12}$ & $-27.2 \,.10^{-12}$ \\
        \hline
    \end{tabular}
    \begin{tabular}{cccc}
        \hline
        \(\chi_{1111}\) & \(\chi_{1123}\) & \(\chi_{1133}\) & \(\chi_{3333}\) \\
        \hline
        $14.07 \,.10^{-21}$ & $3.35 \,.10^{-21}$ & $3.9 \,.10^{-21}$ & $3.35 \,.10^{-21}$ \\
        \hline
    \end{tabular}
    \label{tab:nl_parameters_LiNbO3}
\end{table}

\begin{figure}[ht]
    \centering
    \includegraphics[width=\linewidth]{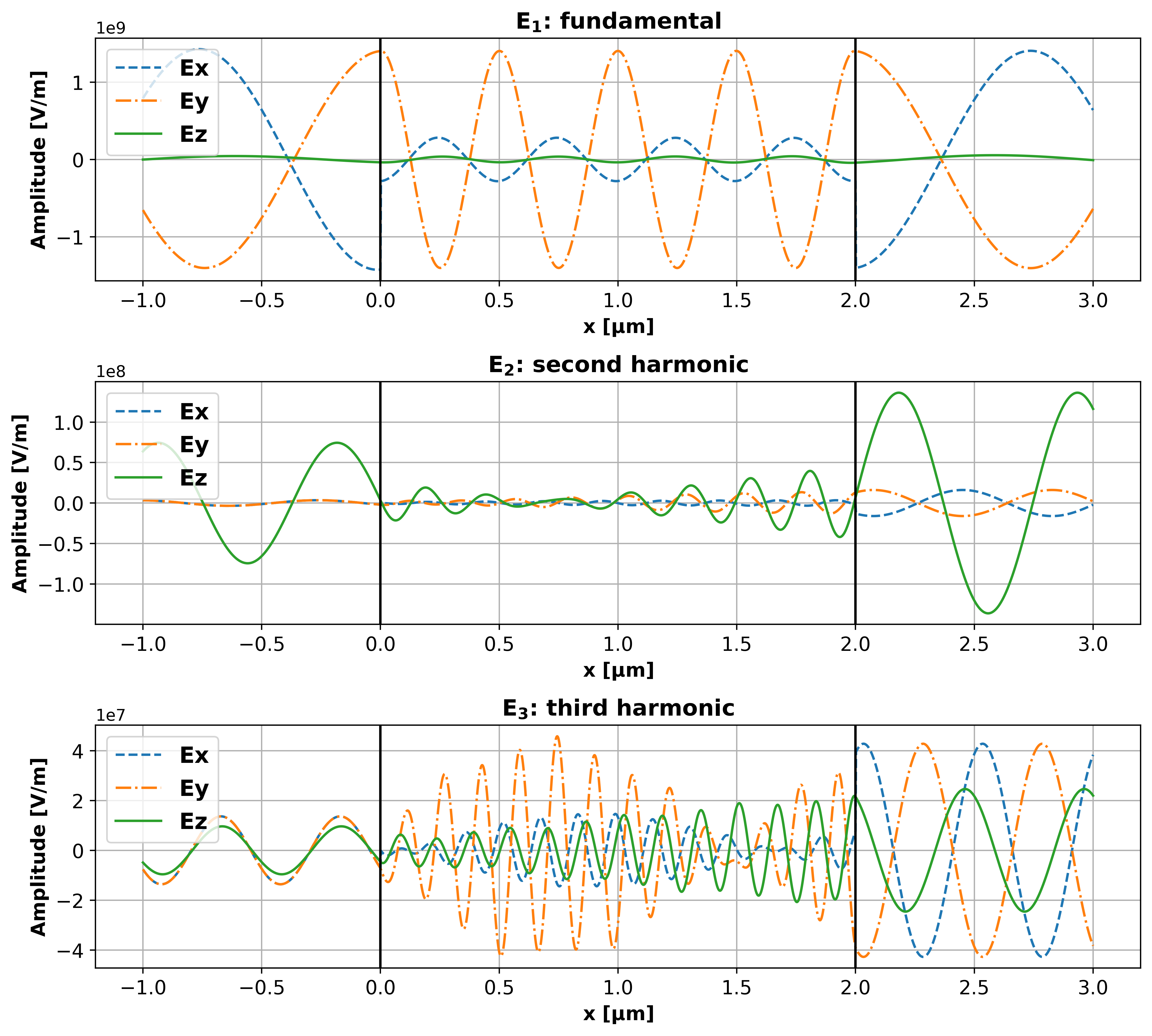}
    \caption{\changecolor{Nonlinear scattering from a LiNbO3 slab under TM-polarized illumination. The real parts of the fundamental field \(\mathbf{E}_1\), the second-harmonic \(\mathbf{E}_2\) and the third harmonic \(\mathbf{E}_3\) are plotted along the \(x\)-axis ($y$$=$$0$), for an incident plane wave from the left with an amplitude \(A_0 = 2\times10^{9} V/m\) and an incidence angle \(\theta=\pi/4 \ rad\).}}
    \label{fig:3HG_TM}
\end{figure}

\changecolor{Fig.~\ref{fig:3HG_TM} shows the real part of the electric field amplitude for an incident TM plane wave. Unlike in the KTP case, all field components are generated because the nonlinear coefficient \(d_{22}\) is nonzero. A significant enhancement of the third harmonic appears near \(0.7\,\mathrm{\mu m}\), while the second harmonic is depleted at the same location, likely due to sum-frequency generation (\(\omega+2\omega\rightarrow3\omega\)). A clear beating of the third harmonic is observed, arising from phase mismatch. We also note a significant backward propagation of the harmonics due to reflections at the slab interfaces. With this configuration, the transmission coefficients \(T_2\) and \(T_3\) are \(0.0047\) and \(0.0011\), respectively, indicating that the third-harmonic generation cannot be neglected in this case. The energy balance has been verified to within \(4.4\times10^{-6}\%\) for a \(10\,\mathrm{nm}\) mesh-size.}

\changecolor{In the next example, we reproduce the setup of Ref.~\cite{szarvas_numerical_2018}, Sec.~3.B. This enables a comparison between our FEM model and a FD-TD model that accounts for pump depletion but only for normally incident waves. We consider a photonic crystal illuminated by a normally incident plane wave of amplitude \(A_0=4\times10^8\,\mathrm{V/m}\); half that in Ref.~\cite{szarvas_numerical_2018} due to a different definition of the field amplitude. The crystal consists of alternating layers of LiNbO3, with one of the two layers rotated by 180° around the x-axis. As a result, both layers share the same permittivity, but their nonlinear coefficients \(\chi_{333}\) have opposite signs. The length \(l\) of the layer is set to be equal to the coherence length of the second-harmonic generation, with \(l=\frac{\lambda_0}{4(n_e^{2\omega}-n_e^{\omega})}=3.406\,\mathrm{\mu m}\).
Index matching is enforced at both ends of the crystal to avoid reflections. Without reflections, backward-propagating waves are negligible compared to forward-propagating ones, so the transmission coefficients can be regarded as the conversion efficiency described in Ref.~\cite{szarvas_numerical_2018}.}

\begin{figure}[ht]
    \centering
    \includegraphics[width=\linewidth]{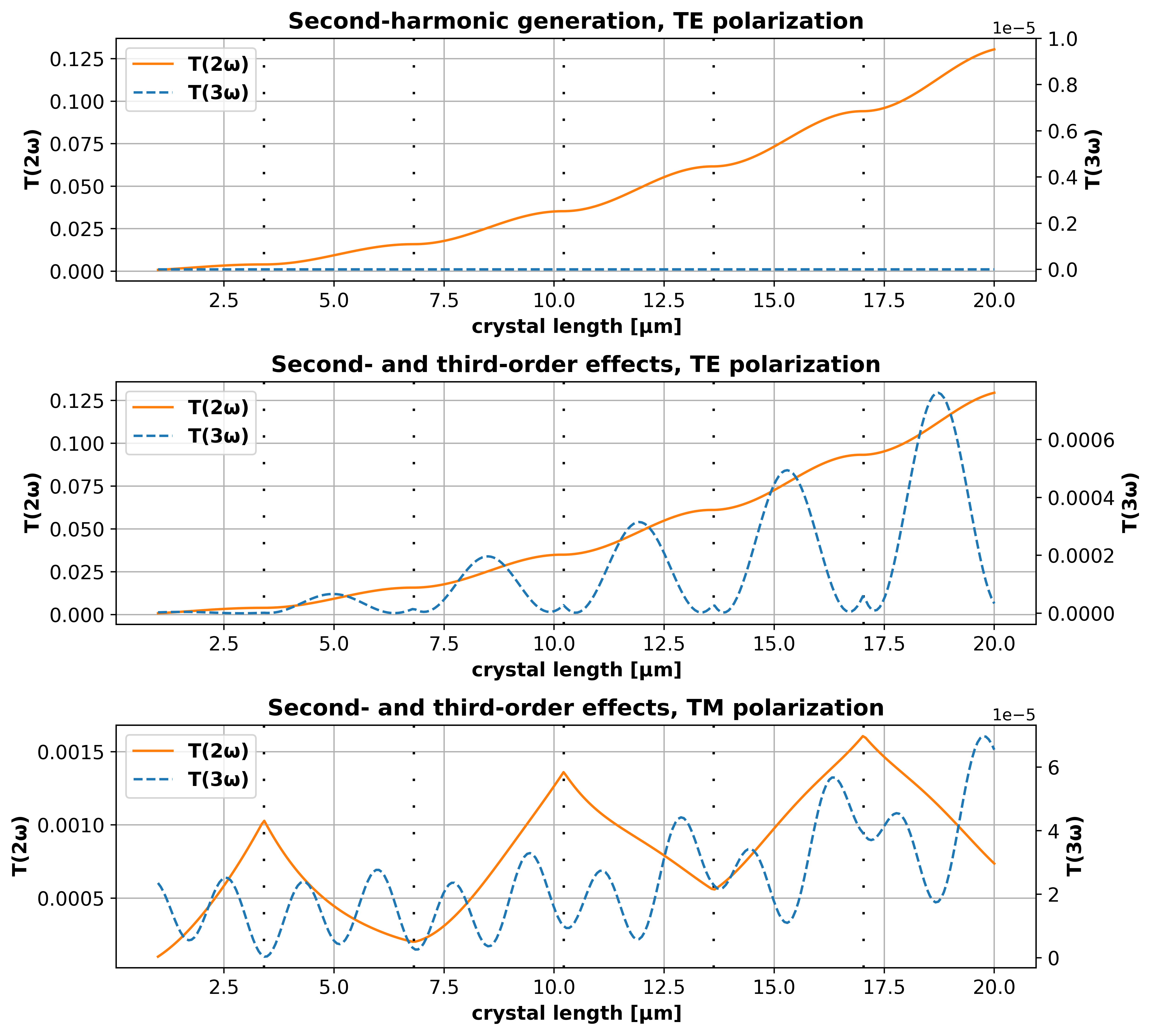}
    \caption{\changecolor{Transmission coefficients \(T_2\) and \(T_3\) as a function of the photonic crystal length. The experiment was carried out for three different configurations: TE polarization including only 2HG and; TE and TM polarization including second- and third-order nonlinear effects. The dotted vertical bars indicate the positions of the layer interfaces.}}
    \label{fig:T_szarvas}
\end{figure}

\changecolor{Fig.\ref{fig:T_szarvas} shows the transmission coefficients \(T_2\) and \(T_3\) of the waves at \(2\omega_I\) and  \(3\omega_I\) as a function of the crystal length for three different cases: (i) a TE-polarized plane wave including only second-harmonic generation, (ii) a TE-polarized plane wave including third-harmonic generation and third-order nonlinearities, and (iii) a TM-polarized plane wave under the same conditions. In the case of second-harmonic generation only, we obtained the same curve as in \cite{szarvas_numerical_2018}, with a transmission coefficient \(T_2\) reaching \(0.125\) for a \(20 \, \mathrm{\mu m}\) crystal. This demonstrates that, for simple configurations (2HG, normal incidence, TE polarization), our model reproduces the results of the nonlinear FD-TD model reported in Ref.~\cite{szarvas_numerical_2018}. When we include third harmonic generation, we observe no significant difference on the energy conversion because of the phase mismatch between the waves at \(\omega_I\) and \(2\omega_I\) with the one at \(3\omega_I\). On the other hand, we observe \(T_3\) oscillating between zero and a curve proportional to \(T_2\). These oscillations arise from the phase mismatch between the refractive indices of the different harmonics. The fact that their maxima scale with \(T_2\) can be attributed to the third harmonic being mainly generated through sum-frequency generation (\(\omega+2\omega\rightarrow3\omega\)) rather than direct third-harmonic generation from the fundamental wave at \(\omega_I\). In the TM case, \(T_2\) and \(T_3\) both exhibit low-amplitude oscillations because the pseudo phase-matching condition is no longer satisfied, since the nonlinear effects now involve all three field components, as shown in Fig.~\ref{fig:3HG_TM}, with six distinct refractive indices and thus six different phase velocities.
The coherence length of the second harmonic should be significantly larger than in the TE case, as the growth of the curve is interrupted at an early stage. It should also be noted that not enough numerical data are reported in~\cite{szarvas_numerical_2018} regarding the accuracy or computation time to allow for a direct comparison with our model and its numerical implementation.}

\subsection{Convergence study}

\begin{figure}[ht]
    \centering
    \includegraphics[width=\linewidth]{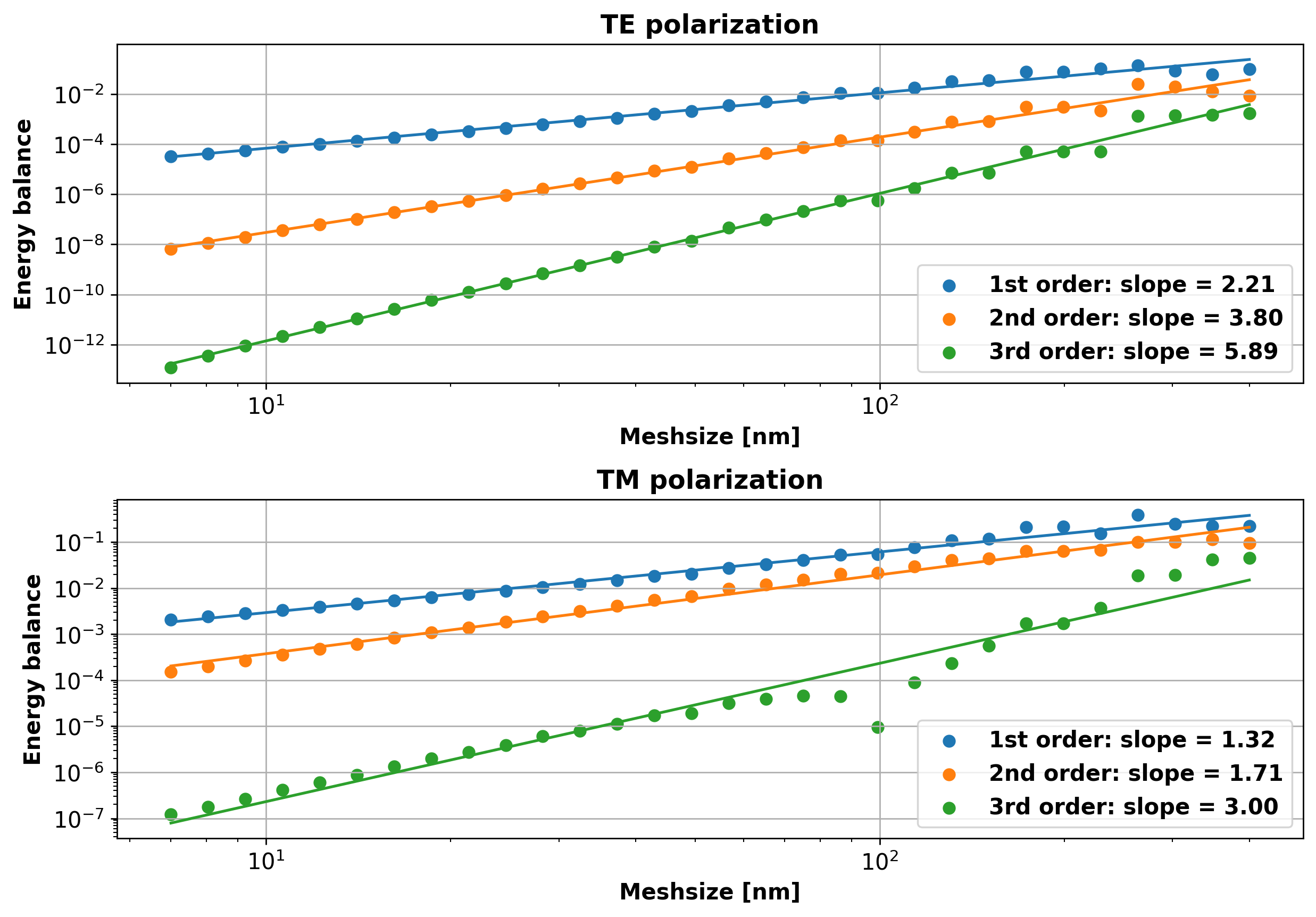}
    \caption{\changecolor{Normalized energy balance \((R_1+T_1+R_2+T_2+Q-1)\) as a function of the mesh size, for first-, second- and third-order FEM elements, in both TE and TM cases, plotted on logarithmic scales.}}
    \label{fig:energyBalance}
\end{figure}

\changecolor{To assess the validity of our model, we investigate its convergence properties in the case of second-harmonic generation, using Eq~\eqref{eq:energy_balance}, as derived earlier. Fig.~\ref{fig:energyBalance} shows the energy balance as a function of the mesh size for first-, second- and third-order FEM elements, under both TE and TM polarization. The material parameters and the incident wave amplitude have been adjusted so that the nonlinear contributions are of similar magnitude in both cases (\(T_2\approx0.25\)): a dispersive slab of thickness \(2 \,\mathrm{\mu m}\) is considered, with \(d_{11}=d_{22}=10^{-12}\,\mathrm{m/V}\), \(d_{33}=6\times 10^{-11}\,\mathrm{m/V}\), illuminated by a plane wave at \(\theta=\pi/6\), with amplitude \(A_0=8\times10^{10}\,\mathrm{V/m}\).}

\changecolor{In the TE case, the slopes are close to 2, 4 and 6, consistent with the use of first-, second- and third-order FEM elements. In contrast, for the TM case, the slopes are noticeably lower. This reduction is attributed to the presence of first derivative terms coupling the \(x\) and \(y\) equations, which reduces the effective accuracy of the simulation. 
We note that by using third-order elements the energy balance reaches \(10^{-12}\) and \(10^{-7}\) for a \(7 \,\mathrm{nm}\) mesh size, in TE and TM polarization, respectively.
The same slopes are recovered for different incident wave amplitudes, albeit at different heights. \emph{Introducing artificial losses through complex refractive indices and nonlinear coefficients does not affect the convergence behavior either.}}

\begin{figure}[H]
    \centering
    \includegraphics[width=\linewidth]{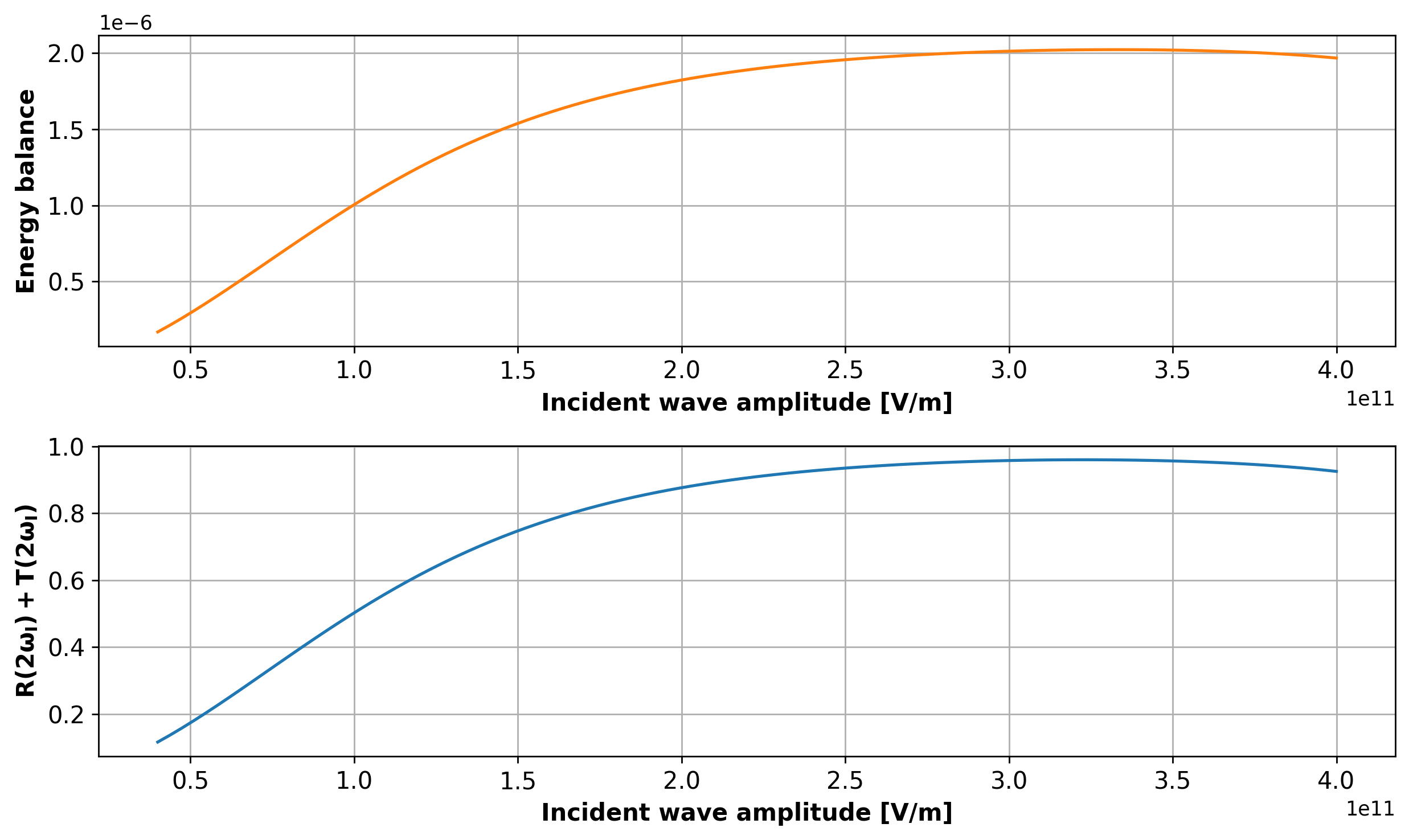}
    \caption{\changecolor{Normalized energy balance \((R_1+T_1+R_2+T_2+Q-1)\) and total second-harmonic energy coefficient \(R_2+T_2\) as functions of the incident wave amplitude, for a TE incident plane wave.}}
    \label{fig:energyBalanceAndEnergyConversion}
\end{figure}

\changecolor{It is also of interest to investigate the impact of nonlinearities on the convergence of the model. Fig.~\ref{fig:energyBalanceAndEnergyConversion} shows the energy balance together with the fraction of energy at \(2\omega_I\), given by \(T_2+R_2\), as a function of the incident wave amplitude. We observe that the error follows the same trend as \(R_2+T_2\): as \(R_2+T_2\) increases, the accuracy decreases. This behavior arises because the wave at \(2\omega_I\) oscillates twice as fast as the wave at \(\omega_I\), thereby requiring twice as many mesh elements to achieve the same level of accuracy.}


\section{Conclusion and outlook}
In line with the work conducted by our research team, we have developed a very general model to simulate the scattering of light by a nonlinear, anisotropic slab, regardless of the incidence angle and polarization. \changecolor{The capabilities of the model are illustrated through two case studies involving two commonly used nonlinear materials, KTP and  LiNbO3, focusing on second-harmonic generation and third-order nonlinear effects,} which clearly demonstrates the expected behavior of energy transfer and phase matching. To validate our approach and test the convergence properties of the numerical model, an energy study was conducted, accounting for possible losses. The simulations reached an accuracy on the order of \(10^{-7}\%\) with a mesh size of \(\lambda_0/100\). While the present paper focuses on second- and third-order nonlinear effects, it is worth noting that the model can also accommodate other types of nonlinearities, involving higher-order susceptibility tensors \(\chi^{(n)}\), as well as the generation of higher harmonics. It should be emphasized that the transition from a two-dimensional problem to a one-dimensional one is only valid for a periodic structure. However, the remainder of this work can be applied to other, more complex, higher-dimensional geometries and sources (as it will be presented in future publications), since the finite element method can handle both two- and three-dimensional problems. We hope that our approach will serve as a valuable tool for overcoming the limitations of existing models and investigating more advanced experimental configurations.

\bibliography{apssamp}% Produces the bibliography via BibTeX.

\end{document}